\documentclass[pra, aps,superscriptaddress, nofootinbib, twocolumn, notitlepage,floatfix,10pt ]{revtex4-2}

\usepackage{amsmath,mathtools,amsthm,amssymb,pifont}
\usepackage[utf8]{inputenc}
\usepackage{graphicx}
\usepackage{ragged2e}
\usepackage{subcaption} 
\usepackage[american]{babel}
\usepackage{graphicx,xcolor,bbold,titlesec}
\usepackage{braket}
\usepackage{MnSymbol}
\usepackage{tabularx}

\usepackage[colorlinks,
bookmarksopen,
bookmarksnumbered,
citecolor=red,
linkcolor=red,
pdfstartview=false,
urlcolor=red]{hyperref}
\usepackage{tikz,ifthen}
\usepackage{bbold}
\usepackage{tikz-network}
\usetikzlibrary{patterns,decorations.pathreplacing,calligraphy}
\usepackage{orcidlink}
\usepackage{color}
\usepackage{MnSymbol}

\begin{document}

\title{Analog quantum simulation of chiral magnetic dynamics using optical superlattices}

\author{Sabhyata Gupta}
\email{sabhyata.gupta@itp.uni-hannover.de}
\affiliation{Institut f\"ur Theoretische Physik, Leibniz Universit\"at Hannover, Appelstr. 2, 30167 Hannover, Germany}

\author{Luis Santos}
\email{santos@itp.uni-hannover.de}
\affiliation{Institut f\"ur Theoretische Physik, Leibniz Universit\"at Hannover, Appelstr. 2, 30167 Hannover, Germany}

\begin{abstract}
We propose an analog quantum simulation of chiral magnetic dynamics using ultracold atoms in an optical superlattice. The massive Schwinger model in the zero gauge coupling limit maps onto the Rice-Mele model, with the fermion mass and topological angle encoded in the superlattice parameters. We study the real-time dynamics of the vector current following two quench protocols that drive continuous chirality injection and chirality relaxation. Simulations with realistic superlattice parameters and experimental noise demonstrates clear mass dependence of the current dynamics in both protocols, robust against experimental imperfections. The vector current may be directly measurable via single-bond-resolved detection, establishing cold atom superlattices as a viable platform for probing non-equilibrium chiral phenomena.
\end{abstract}

\maketitle

\section{Introduction}

The chiral magnetic effect (CME) is a non-equilibrium phenomenon that refers to the generation of an electric current in an external magnetic field in the presence of chiral imbalance \cite{kharzeev2014strongly}. In high-energy physics, the CME has been proposed as a diagnostic of topological charge fluctuations in quark-gluon plasmas \cite{kharzeev2016cme_review}. In condensed matter, the CME manifests in Weyl semi-metals, where nontrivial Berry curvature and chiral node separation give rise to analogous chiral-anomaly-induced transport \cite{SonSpivak2013,WeylAndreev16}. This has enabled controlled tabletop observations of chiral current response, including negative magnetoresistance in various materials~\cite{ZrTe5Li2016, TaAsCMEPhysRevB.101.125102,weyl2025}.

From a theoretical standpoint, the simulation of the dynamics of CME-related transport remains challenging due to sign problems and entanglement growth in none-quilibrium quantum systems. Quantum simulation has emerged as a powerful tool for such problems~\cite{Georgescu2014}. The Schwinger model~\cite{ThetaCOLEMAN1976239}, (1+1)D quantum electrodynamics, provides a minimal gauge theory framework exhibiting chiral anomaly physics and is often used as a toy model for CME dynamics. Digital quantum simulations of the Schwinger model with a topological angle $\theta$-term or with a chiral chemical potential $\mu_5$ have demonstrated basic CME features including anomaly-driven current generation and chirality relaxation~\cite{Kharzeev2020digitalquantumsimulation,FiniteTempCMEDQSIkeda2024}, though system sizes remain small and coherence limited. Both digital approaches based on universal quantum gates and analog approaches where a physical system directly realizes the target Hamiltonian have seen rapid progress~\cite{BlochDalibardNascimbene2012,Daley2022}. In particular ultracold atoms in optical lattices have proven particularly versatile analog simulators, offering precise control over lattice geometry and filling, as well as coherence times far exceeding those of current digital processors~\cite{Gross2017,Altman2021}.

Cold atom platforms offer a promising analog route to simulate anomaly-induced dynamics. Prior proposals have explored the CME in rotating trapped Fermi gases with Weyl-Zeeman spin-orbit coupling~\cite{huang2016soc}, and 
in three-dimensional optical lattices where Weyl semi-metal band structures may be realized using spin-orbital coupling and laser-assisted tunneling~\cite{CME_3D_Optical_lattices_2019}. Quantum link models and staggered fermion constructions have been used to implement aspects of the Schwinger model in one-dimensional cold-atom systems~\cite{schweizer2019gauge,AidelsburgerReview2022,DalmonteMontangero2016,ZoharCiracReznik2016}. Closely related, it has been demonstrated that $\theta$-quenches in the massive Schwinger model produce dynamical quantum phase transitions between topological sectors accessible with cold-atom platforms~\cite{Zache2019}, motivating our analog implementation of the chirality relaxation dynamics studied digitally in Ref.~\cite{Kharzeev2020digitalquantumsimulation}.

In this paper, we propose a concrete and experimentally accessible analog quantum simulation of chiral magnetic dynamics using ultracold atoms in an optical superlattice. The central result is a mapping between the massive Schwinger model in the zero gauge coupling limit and the Rice-Mele tight-binding model, with the fermion mass and topological angle encoded in the superlattice parameters. We discuss in detail the superlattice-based quantum simulation of different non-equilibrium protocols, discussing the effect of systematic errors and possible deviations between the exact evolution and the simulated one. Our results show that superlattices constitute 
a promising platform for the study of CME-related dynamics in state-of-the-art optical lattice experiments.

The paper is organized as follows. In Sec.~\ref{sec:model} we introduce the massive Schwinger model with a time-dependent $\theta$ term, discuss two quench protocols, and present the lattice formulation. In Sec.~\ref{sec:hameng} we discuss the mapping to the Rice-Mele model, and the realization using optical superlattices. In Sec.~\ref{sec:results} we present our results for both quench protocols using realistic superlattice parameters, discussing the effect of systematic errors and possible deviations between the exact evolution and that simulated using superlattices. We conclude in Sec.~\ref{sec:conclusion}.


\section{Theoretical Model}
\label{sec:model}

\subsection{Chiral anomaly and chirality flipping}

In the presence of a strong magnetic field, the dynamics of chiral fermions become effectively $(1+1)$-dimensional, and the chiral magnetic effect can be described within the massive Schwinger model~\cite{ThetaCOLEMAN1976239,kharzeev2016cme_review}.
At the classical level the axial current $J^\mu_5 = \bar{\psi}\gamma^\mu\gamma_5\psi$ is conserved, but upon quantization this symmetry is broken by the chiral anomaly. In $(1+1)$-dimensional QED with massive fermions the anomaly relation takes the form
\begin{equation}
\partial_\mu J^\mu_5 = \frac{1}{\pi}E + 2im\,\bar{\psi}\gamma_5\psi,
\label{eq:anomaly}
\end{equation}
where $E$ is the electric field and $m$ the fermion mass. The first term describes anomalous axial charge nonconservation induced by the gauge field, while the second encodes chirality-flipping processes due to finite fermion mass. In the chiral limit $m = 0$ a chiral quench cannot generate a vector current because the Hamiltonian commutes with the vector current operator~\cite{Kharzeev2020digitalquantumsimulation}. At finite mass this commutator no longer vanishes, allowing a chirally imbalanced state to decay toward the ground state and generate a transient vector current, producing a rich interplay between anomaly-driven chirality pumping and mass-induced relaxation.

In the Schwinger model a background electric field is introduced through a $\theta$-term, $E_\mathrm{cl} = g\theta/2\pi$~\cite{ThetaCOLEMAN1976239}. Applying the chiral rotation $\psi \to e^{i\gamma_5\theta/2}\psi$ absorbs the $\theta$-term into a phase of the fermion mass, and taking $g \to 0$ decouples the gauge dynamics entirely. In this limit a time-dependent $\theta(t)$ induces a chiral imbalance through an effective chiral chemical potential~\cite{ChiralChemicalPotentialOriginKHARZEEV2010205}
\begin{equation}
\mu_5 = -\frac{\dot{\theta}}{2},
\label{eq:mu5}
\end{equation}
At $g = 0$ the electric field vanishes, and hence any vector current generated after the quench is of
purely non-anomalous, mass-driven origin.

\subsection{Quench protocols}
\label{subsec:quench}

To probe the real-time generation and relaxation of chirality, we follow Ref.~\cite{Kharzeev2020digitalquantumsimulation} and consider two spatially uniform
quench protocols that drive the system out of equilibrium from the same initial state, the
ground-state of the unperturbed Hamiltonian at $\theta = 0$. 


\paragraph*{(i) Topological angle quench.} $\theta$ undergoes at $t=0$ 
an instantaneous chiral kick, jumping to a finite value $\theta_f$, remaining constant thereafter. The
induced vector current subsequently relaxes back toward zero, with the relaxation rate and
the presence of oscillations both controlled by the fermion mass $m$~\cite{Kharzeev2020digitalquantumsimulation,Zache2019}.

\paragraph*{(ii) Chiral chemical potential quench.}
At \(t = 0\) a finite chiral chemical potential \(\mu_5\) is switched on and held
constant, corresponding to a linearly growing topological angle \(\theta(t) = -2\mu_5 t\)
for \(t > 0\). Since \(\dot{\theta} = -2\mu_5 = \mathrm{const}\), this protocol sustains
a persistent chirality imbalance between left- and right-handed fermions.

\subsection{Lattice formulation}
\label{subsec:lattice}

We adopt the staggered fermion discretization of the $(1+1)$-dimensional Schwinger model, placing the theory on a spatial lattice of $N$ sites with spacing $a$~\cite{KogutSusskind1975}. 
The two-component Dirac spinor is represented by one-component staggered fermions $\chi_n$, with even (odd) sites carrying the upper (lower) spinor component and hopping amplitude $w = (2a)^{-1}$. We impose open boundary conditions, which avoid the fermion doubling problem and simplify enforcement of the physical subspace~\cite{PhysRevResearch.4.043133}. The staggered Hamiltonian  acquires the form~(see App.~\ref{app:lattice}):
\begin{align}
H &= -iw\sum_{n=0}^{N-2}\!\left(\chi_{n+1}^\dagger\chi_n 
- \mathrm{h.c.}\right) 
- \frac{\dot{\theta}}{4}\sum_{n=0}^{N-2}\!\left(\chi_{n+1}^\dagger\chi_n 
+ \mathrm{h.c.}\right) \notag\\
&\quad + m\cos\theta\sum_{n=0}^{N-1}(-1)^n\chi_n^\dagger\chi_n 
\notag\\
&\quad + \frac{im\sin\theta}{2}\sum_{n=0}^{N-2}(-1)^n\!\left(\chi_{n+1}^\dagger\chi_n 
- \mathrm{h.c.}\right),
\label{eq:latticeH}
\end{align}
where the first line contains the kinetic hopping and axial drive, and the second and the third are the real and imaginary parts of the chirally rotated mass $me^{i\gamma_5\theta}$. This Hamiltonian is the starting point of the present work.

The primary observable of interest is the spatially averaged vector current $\bar{J}$, which is the direct lattice analogue of the CME current. In the staggered fermion representation it takes the form
\begin{equation}
\bar{J} = -\frac{iw}{N}\sum_{n=0}^{N-2}\!\left(\chi_{n+1}^\dagger\chi_n - \mathrm{h.c.}\right),
\label{eq:Jbar}
\end{equation}
corresponding to the continuum bilinear $\bar{\psi}\gamma^1\psi$. It is a nearest-neighbour hopping operator and is gauge-invariant by construction under the $g \to 0$ limit adopted throughout. The time evolution of $\bar{J}(t)$ following each quench protocol directly reflects the interplay between anomaly-driven chirality pumping and mass-induced relaxation, and is the central quantity studied in Sec.~\ref{sec:results}.


\section{Hamiltonian Engineering}
\label{sec:hameng}

\subsection{Mapping to the Rice-Mele model}
\label{sec:mapping}

Hamiltonian \eqref{eq:latticeH} can be written in the form:
\begin{equation}
\!\!H = -\!\sum_{n=0}^{N-2}\!T_n\left(e^{i\Phi_n}\,\chi_{n+1}^\dagger\chi_n + \mathrm{H.c.}\right) + \Delta\sum_{n=0}^{N-1}(-1)^n\,\chi_n^\dagger\chi_n,
\label{eq:H_complex}
\end{equation}
with $\Delta = m\cos\theta$ and 
$T_n\,e^{i\Phi_n} = \frac{\mu_5}{2} 
+ i \left ( w - \frac{m\sin\theta}{2}(-1)^n \right )$. Introducing the gauge transformation $\chi_n = e^{i\beta_n}\,\tilde\chi_n$, 
with $\beta_{n+1} - \beta_n = \Phi_n$,  $\beta_0 = 0$, the Hamiltonian transforms into:
\begin{eqnarray}
H &=& -\sum_{n=0}^{N-2}T_n\left(\tilde\chi_{n+1}^\dagger\tilde\chi_n + \mathrm{H.c.}\right) \nonumber \\&+& \sum_{n=0}^{N-1} \left ( \hbar \dot\beta_n +\Delta(-1)^n \right )\,\tilde\chi_n^\dagger\tilde\chi_n,
\label{eq:H_complex2}
\end{eqnarray}
If 
\begin{equation} 
\frac{m}{2} \ll \sqrt{(\mu_5/2)^2 + w^2},
\label{eq:small_m}
\end{equation}
 we may approximate $T_n \simeq t - (-1)^n\delta$, with $t=\sqrt{\left(\tfrac{\mu_5}{2}\right)^{2} + w^2}$, and $\delta =\frac{wm\sin\theta}{2t}$. Moreover, $\dot \beta_{n\in \mathrm{odd}}=0$, and 
 $\dot\beta_{n\in\mathrm{even}}\simeq\frac{\mu_5^2}{t}\frac{m}{2t}\cos\theta \ll 1$. Hence,  
Eq.~\eqref{eq:H_complex2} reduces to the well-known Rice-Mele Hamiltonian~\cite{RiceMele1982}:
\begin{equation}
\!\!\tilde H\! =\! -\!\sum_{n=0}^{N-2} \!\left (t\!-\!(-1)^n \delta \right )\! \left(\tilde\chi_{n+1}^\dagger\tilde\chi_n \!+\! \mathrm{H.c.}\!\right)\! +\! \Delta\!\sum_{n=0}^{N-1}(-1)^n\,\tilde\chi_n^\dagger\tilde\chi_n.
\label{eq:H_RM}
\end{equation}
The accuracy of the small-mass expansion~\eqref{eq:small_m}, and hence of the mapping to the Rice-Mele model, is discussed in Sec.~\ref{sec:validity}. Note that for the case of the topological angle quench, since $\mu_5 = 0$, the mapping on the Rice-Mele model is exact, without requiring 
condition~\eqref{eq:small_m}, and it is achieved by a simpler gauge transformation 
$\chi_n = e^{i n\pi/2} \tilde\chi_n$. 



\begin{figure}[htbp]
    \centering
    \includegraphics[width=0.8\columnwidth]{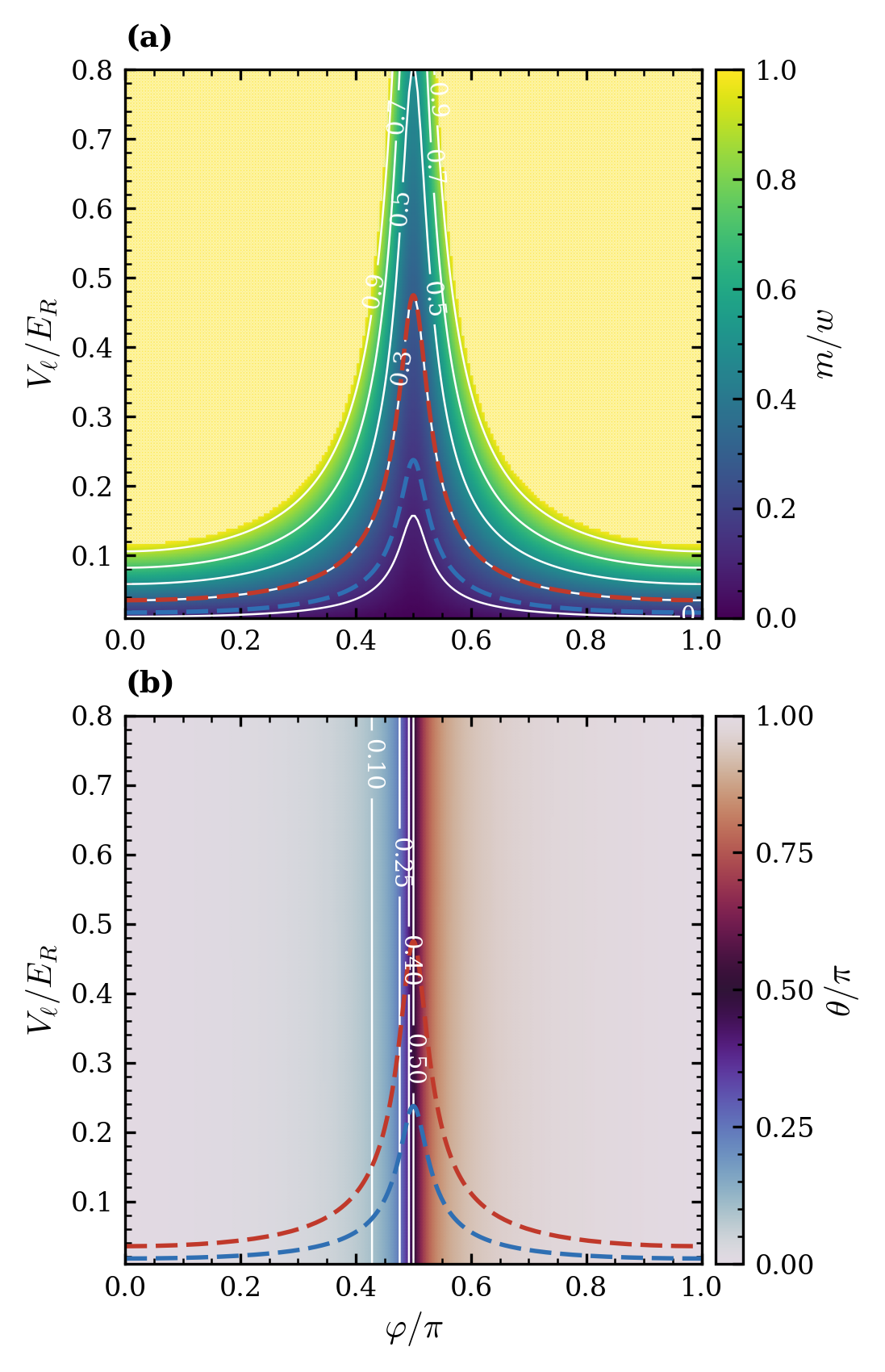}
    \caption{\justifying
    Realized mass $m$~(a) and topological angle $\theta$~(b) as a function of the relative phase $\varphi$ and depth $V_\ell$ of the secondary lattice, at fixed primary-lattice depth $V_s = 8\,E_R$, with $E_R$ the recoil energy. 
    White curves are contours of constant $m/w$ and $\theta/\pi$. Dashed lines trace the curves  $V_\ell(\varphi)$ that correspond to a fixed $m/w=0.15$ (blue) and $m/w=0.30$ (red), while $\theta$ is varied.
    In (a) the colormap saturates for $m/w>1$, since that region lies outside the light-mass regime of interest.}
    \label{fig:parammap}
\end{figure}


The vector current~\eqref{eq:Jbar} acquires the form:
\begin{equation}
\bar J = -\frac{w}{N}\sum_n \left ( 
\cos\Phi_n \, \hat\kappa_n + \sin\Phi_n\, \hat\beta_n 
\right ),
\label{eq:J_split}
\end{equation}
where $\hat\kappa_n = i \left (\tilde\chi_{n+1}^\dagger\tilde\chi_n -\mathrm{H.c.}\right)$ is the local current, and 
$\hat\beta_n = \left (\tilde\chi_{n+1}^\dagger\tilde\chi_n +\mathrm{H.c.}\right)$ is associated to the kinetic energy. Both $\langle \hat\kappa_n \rangle$ and 
$\langle \hat \lambda_n \rangle$, and hence $\bar J$, may be experimentally determined by means of recently-developed dimerization techniques~\cite{AidelsburgerPhysRevLett.133.063401}.


\subsection{Optical superlattice realization}
\label{sec:superlattice}

The Rice-Mele model may be realized using ultracold atoms in an optical superlattice. This platform allows for a highly tunable manipulation of both $\Delta$ and $\delta$ as shown in recent experiments on Thouless pumping~\cite{Lohse2016,Walter2023,Viebahn2024}. The lattice potential:
\begin{equation}
V(x)=V_s \cos^2(\pi x) + V_\ell \cos^2(\pi x/2 + \varphi/2),
\end{equation}
is composed by a primary lattice of depth $V_s$ (with lattice constant $\lambda=1$) and a secondary lattice of half the wavevector, depth $V_\ell$, and relative phase $\varphi$, which generates the sublattice offset $\Delta$ and the bond dimerization $\delta$. 
Figure~\ref{fig:parammap} shows the mapping of $(V_\ell,\varphi)$ onto $(m,\theta)$ obtained from the superlattice band structure, see App.~\ref{app:bandstructure}. The topological angle is governed almost entirely by $\varphi$, as seen from the nearly vertical contours in Fig.~\ref{fig:parammap}(b), while the mass is set by $V_\ell$, scaling approximately linearly with it. A target $(m,\theta)$ is thus reached by choosing $\varphi$ to fix $\theta$ and $V_\ell$ to fix $m$; holding the mass constant while scanning the angle traces the schedule $V_\ell(\varphi)$,  shown as dashed curves for $m/w = 0.15$ and $0.30$. The topological angle spans the interval $\theta \in [0,\pi)$, and the mass is tunable up to $m \sim w$. 
We recall, however, that for the 
chiral chemical potential quench, since $\mu_5\neq 0$, the mapping to the Rice-Mele Hamiltonian demands the
condition~\eqref{eq:small_m}, and hence 
quantum simulations 
for larger $m/w$ are expected to deviate from the actual model~\eqref{eq:latticeH}. We discuss this point in more detail in Sec.~\ref{sec:validity}.

The envisaged quantum simulation starts with the initial state preparation, where the lowest lattice band (at $\theta=0$) is adiabatic loaded at half-filling. The $\theta$ quench is achieved by a sudden jump in $\varphi$, e.g. set by an electro-optic modulator, on timescales short compared to the tunneling time, $1/t$~\cite{Walter2023,Viebahn2024}. This jump in $\varphi$ realizes the targeted $\theta=\theta_f$ at fixed mass. The $\mu_5$ quench demands a coordinated continuous ramp of both $V_\ell$ and $\varphi$. This protocol is the more demanding, as holding the mass fixed while rotating $\theta$ requires a sizeable variation of $V_\ell$. With the demonstrated tunneling rate $t/h \simeq 370$\,Hz~\cite{Viebahn2024}, our typical simulation time $\tau< 6/t$ is well within the range over which quantized transport is sustained in Ref.~\cite{Viebahn2024}, which exceeds one hundred tunneling times. The complete transient dynamics hence unfolds comfortably within the coherence window. 
Both quench protocols are spatially uniform and require no single-site addressing.



\begin{figure}[t]
    \centering
    \includegraphics[width=0.8\columnwidth]{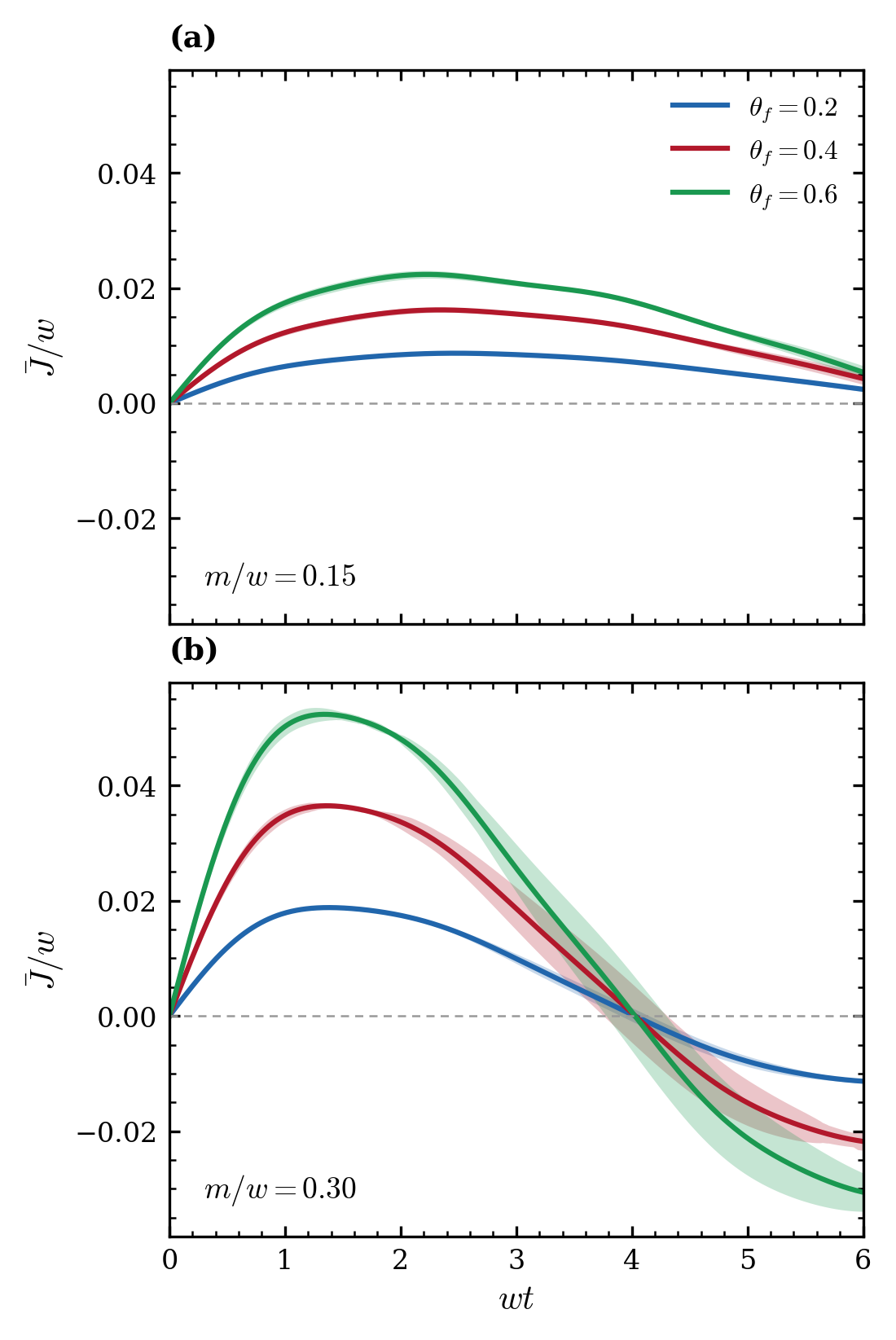}
    \caption{\justifying
    Vector current $\bar J$ following a $\theta$ quench from $\theta=0$ to $\theta_f$. The time evolution is computed using the Rice-Mele Hamiltonian on an open
    chain of $N=30$ sites, and $\bar{J}$ is determined using Eq.~\eqref{eq:J_split}. Panels show
    (a)~$m/w=0.15$ and (b)~$m/w=0.30$, with $\theta_f=0.2$~(blue), $0.4$~(red),
    and $0.6$~(green). Solid lines are the central (unperturbed) configuration, whereas 
    shaded regions are bands obtained from systematic perturbations of the
    superlattice parameters $V_s$~($\pm2.5\%$), $V_\ell$~($\pm2.5\%$), and
    $\varphi$~($\pm0.02$ rad). The dashed region is given by the maximal and minimal value obtained within those error windows.}
    \label{fig:theta_quench}
\end{figure}




\begin{figure}[t]
    \centering
    \includegraphics[width=0.8\columnwidth]{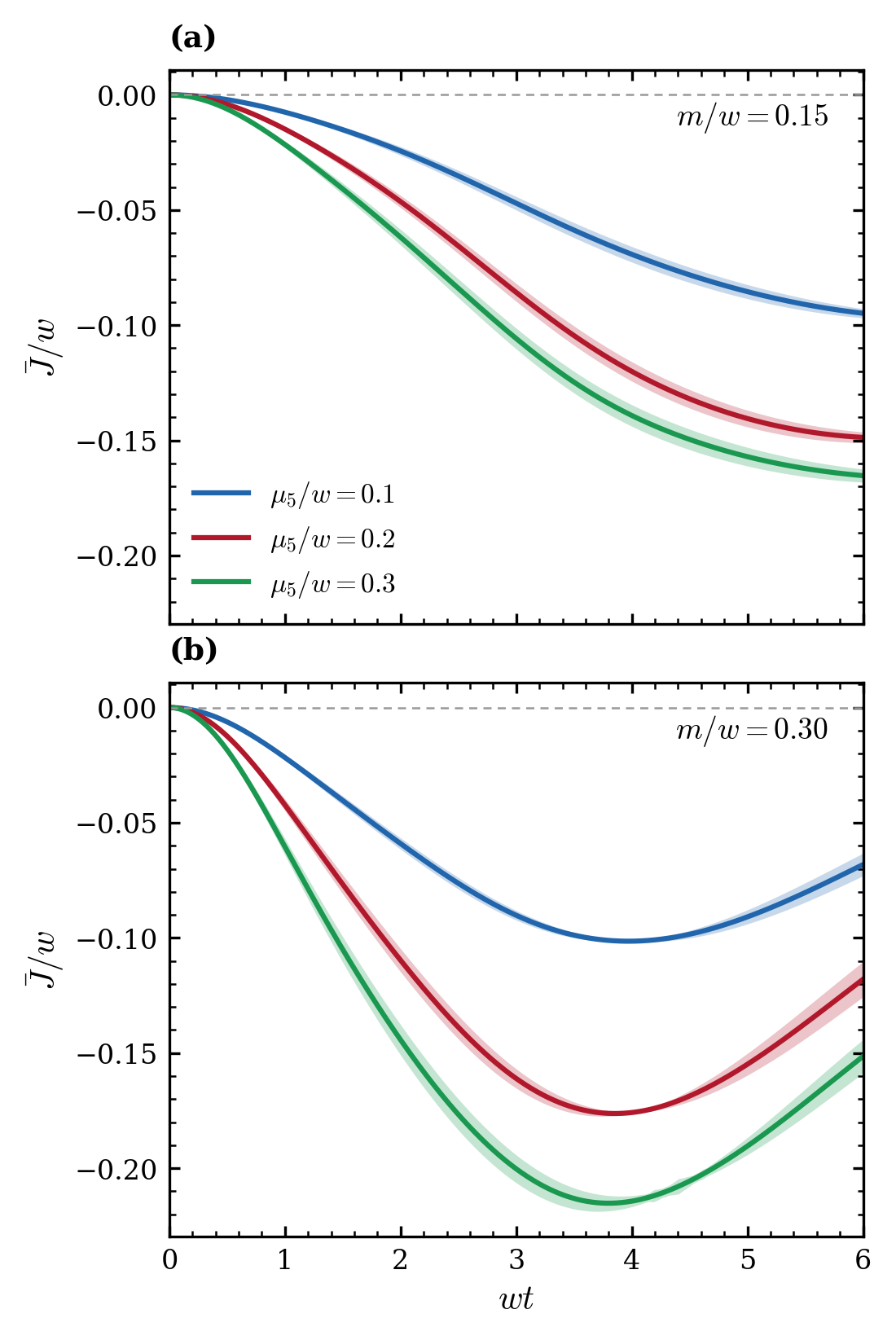}
    \caption{\justifying
    Vector current $\bar J$ following a $\mu_5$ quench, $\theta(t)=-2\mu_5 t$. The evolution is computed using the Rice-Mele Hamiltonian on an open chain of $N=30$ sites. Panels show (a)~$m/w=0.15$ and
    (b)~$m/w=0.30$, with $\mu_5/w=0.1$~(blue), $0.2$~(red), and $0.3$~(green).
    Solid lines and shaded regions are defined as in Fig.~\ref{fig:theta_quench}.}
    \label{fig:mu5_quench}
\end{figure}


\section{Results}
\label{sec:results}

In the following,  we present our numerical results of the evolution of the vector current $\bar J$. In all our simulations, we start at $\theta=0$~($\delta=0$, $\Delta=m$), employing exact numerical evolution on an open chain of $N=30$ sites. 
The current $\bar J$ is evaluated as an average over the interior bonds, excluding three sites at each edge to suppress boundary effects. 

\subsection{Topological angle quench}

Figure~\ref{fig:theta_quench} shows for $m/w=0.15$ and $m/w=0.30$, the current $\bar J$ after a sudden change of $\theta$ from $0$ to $\theta_f$. We recall that for the $\theta$ quench the mapping to the Rice-Mele model is exact, and hence the quantum simulation of model~\eqref{eq:latticeH} is in principle exact.
However, experimentally, there may be systematic errors 
in the superlattice parameters. The shaded bands in Figs.~\ref{fig:theta_quench} are obtained by considering 
$\pm2.5\%$ deviations in the value of $V_s$ and $V_\ell$, and a deviation of $\pm0.02$\,rad in the value of $\varphi$. The systematic errors introduce a non-negligible effect at longer times, especially for larger masses and larger $\theta_f$ values,
due to the stiff dependence of $\theta$ on $\varphi$, see Fig.~\ref{fig:parammap}~(b). 
The errors, however, do not compromise the visibility of the overall dynamics.

Starting from zero, the current rises, reaches a maximum, and relaxes back toward zero, reflecting the transient chirality pumping and subsequent mass-induced relaxation that follow the quench. The peak amplitude grows with both $\theta_f$ and the mass. For $m/w=0.30$ the current changes sign at late times, a direct signature of the faster mass-induced chirality relaxation.

\subsection{Chiral chemical potential quench}

Figure~\ref{fig:mu5_quench} shows our results for 
the $\mu_5$ quench, for which $\theta(t)=-2\mu_5 t$, for a constant $\mu_5$. We depict the case of different $\mu_5$ values, and two different masses $m/w=0.15$ and $0.30$. We recall that the map to the Rice-Mele model is not exact for the $\mu_5$-quench. We will discuss this point in more detail in the next subsection. 

The current rises again from zero but, in contrast to the $\theta$ quench, it is continuously driven. It builds up and tends toward a finite value set by the balance between the injection rate $\mu_5$ and the mass-induced relaxation. The current scales with $\mu_5$ and, at the larger mass panel~(b), saturates earlier as relaxation becomes more effective. 
At short times the growth is quadratic, consistent with the perturbative result $\bar J \propto -m^2 t^2$~\cite{Kharzeev2020digitalquantumsimulation}. This is the regime directly relevant to the real-time chiral magnetic effect.

The effect of the systematic errors discussed above is again indicated with shaded areas. Our results confirm that the mass dependence of the chiral dynamics is observable under realistic laser-power and phase fluctuations.  Unlike the $\theta$ quench, for the $\mu_5$ quench the effect of fluctuations is significantly smaller. This is because $\theta(t)=-2\mu_5 t$, and hence the phase error, which as mentioned above is larger for larger $\theta$, is averaged over the whole angle sweep, suppressing to a large extend the systematic error.



\begin{figure}[htbp]
  \centering
  \includegraphics[width=0.8\columnwidth]{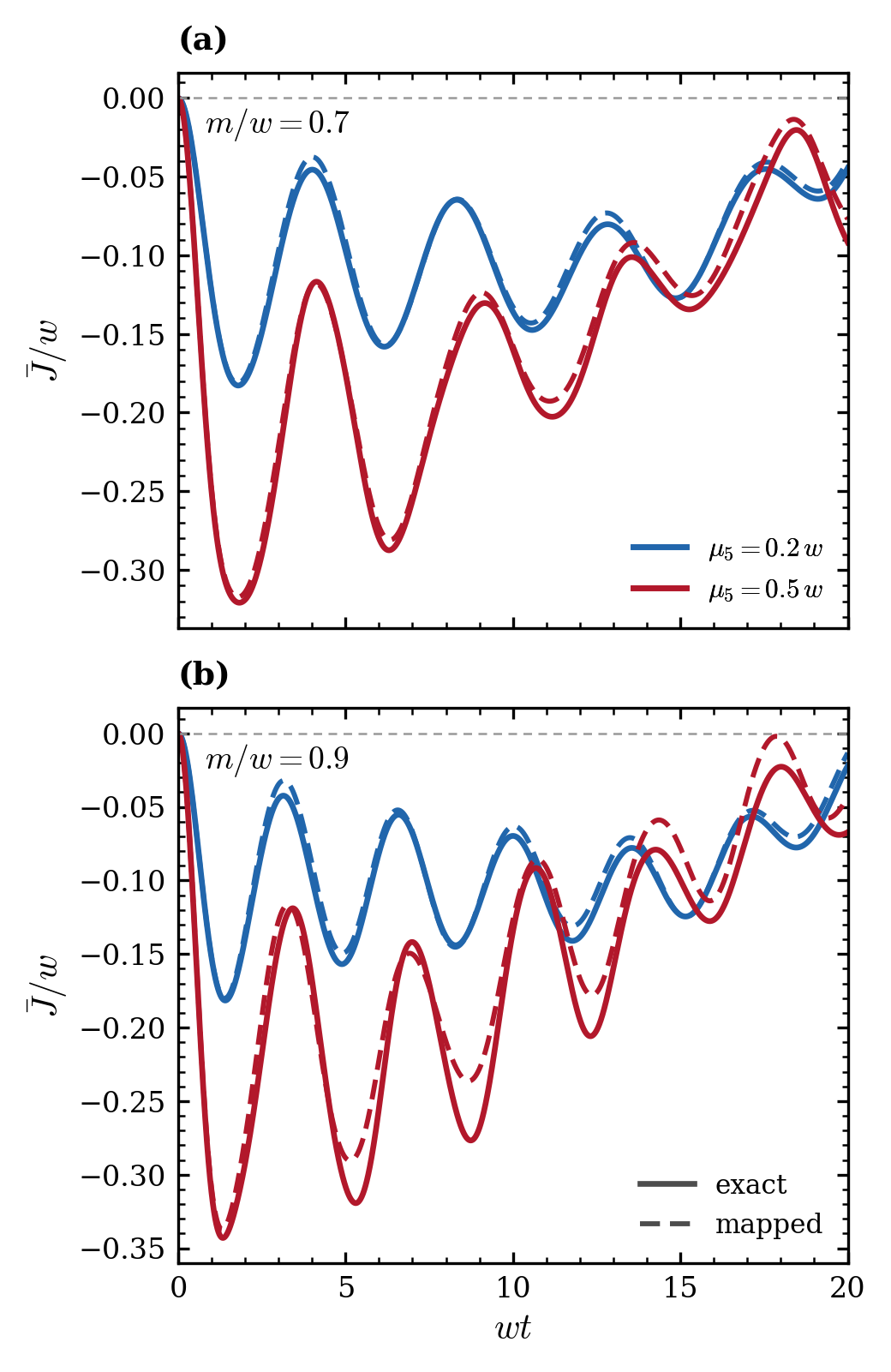}
  \caption{\justifying Validity of the superlattice-based quantum simulation for the $\mu_5$ quench for (a) $m/w=0.7$ and (b) $m/w=0.9$. The
  vector current $\bar J/w$ evaluated using the Rice-Mele model~(dashed curves) is compared with the exact dynamics given by model~\eqref{eq:latticeH}~(solid curves), for $\mu_5=0.2\,w$~(blue) and $\mu_5=0.5\,w$~(red). }
  \label{fig:validity}
\end{figure}


\subsection{Validity of the superlattice simulation}
\label{sec:validity}

The mapping onto the Rice-Mele model 
relies, for the case of the $\mu_5$ quench, on the condition~\eqref{eq:small_m}. The accuracy of the superlattice-based quantum simulation of model~\eqref{eq:latticeH}  is hence expected to degrade as $m/w$ grows.
Figure~\ref{fig:validity} quantifies this by comparing 
the results for the time evolution of $\bar{J}$ 
evaluated using Eq.~\eqref{eq:latticeH}~(solid lines) and using the Rice-Mele model~\eqref{eq:H_RM}. 
We consider two relatively large values $m/w=0.7$ and $m=0.9$~(for the low values of $m/w$ employed in Fig.~\eqref{fig:mu5_quench}, the agreement is basically exact).
For moderate $\mu_5=0.2\,w$ the superlattice-based quantum simulation is expected to mirror the exact result almost perfectly, with only minor deviations developing at late times. For the faster drive $\mu_5=0.5\,w$ the current observed in the quantum simulator departs visibly from the exact result as time progresses. As expected, the departure grows when $m/w$ increases. 

The superlattice-based quantum simulation is hence expected to remain faithful up to at least $20$ hopping times in the regime of moderate mass $m/w\lesssim 0.5$ and driving $\mu/w\lesssim 0.5$, as that considered in Figs.~\ref{fig:theta_quench} and~\ref{fig:mu5_quench}.


\section{Conclusion and outlook}
\label{sec:conclusion}

Experiments with optical superlattice provides a controllable platform for the analog quantum simulation 
of the real-time chiral dynamics of the massive Schwinger model. We have shown that, to a good approximation, the staggered Schwinger Hamiltonian maps onto the Rice-Mele model, with the fermion mass $m$ and topological angle $\theta$ encoded in the parameters of the secondary lattice. We have studied this superlattice-based quantum simulation for the case of a sudden quench of $\theta$, and for a continuous injection $\theta = -2\mu_5 t$. We have shown that the dynamics of the generated vector current, and its dependence on mass and quench parameter, may be faithfully simulated, being clearly resolved and robust against realistic laser-power and phase fluctuations. The full transient unfolds within a few tunneling times which are well inside the demonstrated coherence window of current superlattice experiments~\cite{Viebahn2024}.

The present work establishes the feasibility of cold atom platforms for studying nonequilibrium chiral phenomena and opens several directions for future work. Beyond the non-interacting dynamics studied here, including the dynamical gauge field would enable simulation of the full Schwinger model~\cite{Schwinger1962} with confinement and string breaking~\cite{Martinez2016}. Finite temperature effects, relevant for comparison with heavy-ion phenomenology~\cite{kharzeev2016cme_review}, can be incorporated by preparing thermal initial states.
Furthermore, the Rice-Mele Hamiltonian describes as well quasiparticles in one-dimensional Dirac and Weyl semimetals~\cite{weyl2025,WeylAndreev16}, so the superlattice platform also constitutes a possible simulator of chiral materials~\cite{ZrTe5Li2016, TaAsCMEPhysRevB.101.125102,weyl2025}. 

\begin{acknowledgments}
S.G. and L.S. acknowledge the support of the Deutsche Forschungsgemeinschaft (DFG, German Research Foundation) under Germany's Excellence Strategy -- EXC-2123 Quantum-Frontiers -- 390837967.
\end{acknowledgments}

\bibliographystyle{apsrev4-2}
\bibliography{references}
\appendix

\section{Lattice formulation of the Schwinger model}
\label{app:lattice}

We recall in this appendix, the derivation of the lattice model~\eqref{eq:latticeH}. For more details we refer e.g. to Ref.~\cite{Kharzeev2020digitalquantumsimulation}. The Lagrangian density of the massive Schwinger model with a time-dependent topological angle $\theta(t)$ is
\begin{equation}
\mathcal{L} = -\frac{1}{4}F_{\mu\nu}F^{\mu\nu} + \frac{g\theta}{4\pi}\epsilon^{\mu\nu}F_{\mu\nu} + \bar{\psi}\left(i\gamma^\mu D_\mu - m\right)\psi,
\label{eq:lagrangian}
\end{equation}
with $D_\mu = \partial_\mu - igA_\mu$. Working in temporal gauge $A_0 = 0$, the canonical momentum conjugate to $A_1$ is $\Pi = \dot{A}_1 - g\theta/2\pi$, and the shift $g\theta/2\pi$ is identified as the classical background electric field $E_\mathrm{cl} = g\theta/2\pi$. Applying the chiral rotation $\psi \to e^{i\gamma_5\theta/2}\psi$ absorbs the $\theta$-term into a phase of the fermion mass and generates the axial bias $\mu_5 = -\dot{\theta}/2$ of Eq.~\eqref{eq:mu5}. Taking $g \to 0$ decouples the gauge sector entirely, leaving the continuum Hamiltonian
\begin{equation}
H = \int dx\,\bar{\psi}\left[\gamma^1\!\left(i\partial_1 - \frac{\dot{\theta}}{2}\right) + me^{i\gamma_5\theta}\right]\psi.
\label{eq:freeH}
\end{equation}
To discretize, the theory is placed on a spatial lattice of $N$ sites with spacing $a$ and represent the Dirac spinor by one-component staggered fermions $\chi_n$, with even (odd) sites carrying the upper (lower) spinor component and hopping amplitude $w = (2a)^{-1}$. The gamma matrices in $(1+1)$ dimensions are chosen as $\gamma^0 = \sigma_z$, $\gamma^1 = i\sigma_y$, $\gamma_5 = \gamma^0\gamma^1 = -\sigma_x$, consistent with the staggered fermion representation. The continuum bilinears map onto staggered operators as
\begin{align}
\int dx\,\bar{\psi}\gamma^1\psi &\to -\frac{1}{2}\sum_n\!\left(\chi_{n+1}^\dagger\chi_n + \mathrm{h.c.}\right), \\
\int dx\,\bar{\psi}\psi &\to \sum_n(-1)^n\chi_n^\dagger\chi_n, \\
\int dx\,\bar{\psi}\gamma_5\psi &\to -\frac{1}{2}\sum_n(-1)^n\!\left(\chi_{n+1}^\dagger\chi_n - \mathrm{h.c.}\right).
\end{align}
Substituting into Eq.~\eqref{eq:freeH} and expanding $me^{i\gamma_5\theta} = m\cos\theta + im\sin\theta\,\gamma_5$ yields the staggered Hamiltonian of Eq.~\eqref{eq:latticeH}. Open boundary conditions are imposed by restricting all link sums to $n = 0, \ldots, N-2$.

\section{Extraction of the coefficients of Model~\eqref{eq:latticeH} from the superlattice potential}
\label{app:bandstructure}

The coefficients of Model~\eqref{eq:latticeH} simulated by the superlattice platform may be 
easily determined from the eigenspectrum of a particle with quasi-momentum $k$ in the super-lattice potential 
$V(x)=V_s \cos^2(\pi x) + V_\ell \cos^2(\pi x/2 + \varphi/2)$, with $\lambda=1$ the lattice constant. In the tight-binding regime, for $V_s/E_R\gtrsim 5$~(with $E_R=\hbar^2\pi^2/2M\lambda^2$ the recoil energy), the lowest two bands match very well with the eigenenergies of the Rice-Mele model, $E_\pm(k)=\pm\sqrt{F(k)}$, with
$F(k) = \Delta^2 + 2(t^2+\delta^2) + 2(t^2-\delta^2)\cos(k)$. We may then easily obtain $t$, $\Delta$ and $\delta$ for a given $V_s$, $V_\ell$ and $\varphi$. From the Rice-Mele coefficients, we may then obtain for a given choice of $\mu_5$:
$w/t \simeq \sqrt{1-\mu_5^2/4t^2}$, 
$m^2 = \Delta^2 + 4\delta^2 (t/w)^2$, and 
$\tan\theta = 2 (\delta/\Delta) (t/w)$.

\end{document}